\documentclass[a4paper,usenatbib]{mnras} 

\usepackage{mathptmx}
\usepackage{xspace}
\usepackage{fleqn}

\usepackage[T1]{fontenc}
\usepackage{ae,aecompl}

\usepackage[draft,textsize=footnotesize]{todonotes}%
\usepackage[normalem]{ulem}%

\usepackage{graphicx}	
\usepackage{amsmath}	
\usepackage{amssymb}	

\usepackage{aasmacros}
\usepackage{units}

\newcommand{\Ls}{$L_\rmn{6}$\xspace}
\newcommand{\Ss}{$\sin^{2}{2\theta}$\xspace}
\newcommand{\Msun}{\mathrm{M_{\sun}}}

\newcommand{\kms}{$\rmn{km\,s^{-1}}$} 
\newcommand{\vmax}{$V_{\rmn{max}}$\xspace}

\def\gsim{ \lower .75ex \hbox{$\sim$} \llap{\raise .27ex \hbox{$>$}} }
\def\lsim{ \lower .75ex \hbox{$\sim$} \llap{\raise .27ex \hbox{$<$}} }
\def\LCDM{$\Lambda${\sc cdm}}
\def\jcap{JCAP}

\title[APOSTLE with sterile neutrinos]{Properties of Local Group galaxies in hydrodynamical simulations of sterile neutrino dark matter cosmologies}
\author[M. R. Lovell et al.]{Mark R. Lovell$^{1,2,3}$\thanks{E-mail: lovell@mpia-hd.mpg.de}, Sownak Bose$^{4}$, Alexey Boyarsky$^{2}$, Robert A. Crain$^{5}$,\newauthor Carlos S. Frenk$^{4}$, Wojciech A. Hellwing$^{6,7}$, Aaron D. Ludlow$^{4}$, Julio F. Navarro$^{8}$,\newauthor Oleg Ruchayskiy$^{9}$, Till Sawala$^{10}$, Matthieu Schaller$^{4}$, Joop Schaye$^{11}$, \newauthor and Tom Theuns$^{4}$\\
    $^{1}$GRAPPA, Universiteit van Amsterdam, Science Park 904, NL-1098 XH
Amsterdam, the Netherlands\\
$^{2}$Instituut-Lorentz for Theoretical Physics, Niels Bohrweg 2, NL-2333 CA Leiden, the Netherlands\\
$^{3}$Max-Planck-Institut f\"ur Astronomie, K\"onigstuhl 17, D-69117 Heidelberg, Germany\\
$^{4}$Institute for Computational Cosmology, Durham University, South Road, Durham, DH1 3LE, UK\\
$^{5}$Astrophysics Research Institute, Liverpool John Moores University, 146 Brownlow Hill, Liverpool L3 5RF, UK\\
$^{6}$Institute of Cosmology and Gravitation, University of Portsmouth, Dennis Sciama Building, Portsmouth PO1 3FX, UK\\
$^{7}$Janusz Gil Institute of Astronomy, University of Zielona G\'ora, ul. Szafrana 2, 65-516 Zielona G\'ora, Poland\\
$^{8}$Department of Physics and Astronomy, University of Victoria, 3800 Finnerty Road, Victoria, BC V8P 5C2, Canada\\
$^{9}$Discovery Center, Niels Bohr Institute, Blegdamsvej 17, 2100, Copenhagen, Denmark\\
$^{10}$Department of Physics, University of Helsinki, Gustaf H\"allstr\"omin katu 2a, FI-00014 Helsinki, Finland\\
$^{11}$Leiden Observatory, Leiden University, PO Box 9513, NL-2300 RA Leiden, the Netherlands\\}

  \date{Accepted *** Received ***; in original
    form ***} 

\begin{document}

  \pagerange{\pageref{firstpage}--\pageref{lastpage}} \pubyear{2016}

  \maketitle

  \label{first page}

  \begin{abstract}
     
 \noindent We study galaxy formation in sterile neutrino dark matter
models that differ significantly from both cold and from
`warm thermal relic' models. We use the {\sc eagle} code to carry
out hydrodynamic simulations of the evolution of pairs of
galaxies chosen to resemble the Local Group, as part of the
APOSTLE simulations project. We compare cold dark matter
(CDM) with two sterile neutrino models with 7 keV mass:
one, the warmest among all models of this
mass (LA120) and the other, a relatively cold case (LA10). We show that the lower concentration of
sterile neutrino subhaloes compared to their CDM counterparts
makes the inferred inner dark matter content of galaxies
like Fornax (or Magellanic Clouds) less of an outlier in the
sterile neutrino cosmologies.  In terms of the galaxy number
counts the LA10 simulations are \emph{indistinguishable} from
CDM when one takes into account halo-to-halo (or
`simulation-to-simulation') scatter.  In order for the LA120
model to match the number of Local Group dwarf galaxies, a
higher fraction of lowf
mass haloes is required to form galaxies than is predicted by the {\sc eagle} simulations.  As the census of
the Local Group galaxies nears completion, this population may provide a strong discriminant between cold and warm dark matter models.
  
  \end{abstract}

  \begin{keywords}
	Local Group -- dark matter
  \end{keywords}

  \section{Introduction}
  \label{intro}

 The identity of the dark matter is a major outstanding problem in physics. If the dark matter consists of some as yet unknown particles, little is known about their properties. The cosmic microwave background measurements convincingly demonstrate that these particles did not couple with the baryon--photon plasma and experienced gravitational clustering well before the recombination epoch \citep[e.g.][]{PlanckCP13}. 
The large-scale structure of the Universe tells us that these particles should have a free-streaming scale smaller than the scale of bright galaxies \citep{Davis85}, and should be collisionless on scales greater than $\sim1$~kpc \citep{Yoshida00}. These constraints are consistent with what one would expect from a weakly interacting massive particle \citep[WIMP;][]{Lee:1977ua}.
Such particles would be copiously produced at Large Hadron Collider (LHC), but
experiments have found no traces of any so far \citep[see e.g.][]{Zerwas:2015gec}. Additionally, attempts to detect weakly interacting dark matter particles  directly have yet to see a signal, despite beginning to probe the weak interaction cross-section \citep{Xenon2012,Lux2014}.
  
 By contrast, astronomical observations have provided some tantalising, if strongly disputed, leads. The {\emph Fermi}-LAT $\gamma$-ray observatory has detected an excess of $\gamma$-ray photons in the Galactic Centre relative to the expected background from astrophysical sources -- the so-called `GeV excess' --  and it has been suggested that this emission is consistent with WIMP dark matter annihilating into various decay products \citep{Hooper11,Abazajian12,Abazajian14b,Calore:2014xka,Fermi15}. Another claimed signal is an unidentified X-ray line feature observed at an energy of 3.55~keV in galaxy clusters \citep{Boyarsky14a,Bulbul14,Iakubovskyi15,Urban:2014yda,Franse16}, M31  \mbox{\citep{Boyarsky14a}}, the Galactic Centre \mbox{\citep{Boyarsky15,Jeltema15}} and the Cosmic X-ray background~ \mbox{\citep{Cappelluti17}}. This line is consistent with the decay of a 7.1~keV sterile neutrino, which is an alternative dark matter candidate that is well motivated on particle physics grounds  \citep{Dodelson94, Shi99, Asaka05, Laine08, Boyarsky09a}. 
 
 Whilst the existence of both the GeV excess and the 3.5~keV line is well established, their dark matter origin hypotheses remain controversial. The slope of the radial profile of the GeV excess is possibly too steep to result from dark matter annihilation \mbox{\citep{Calore15,Schaller16}} and the signal properties can be readily explained by a population of unresolved millisecond pulsars \citep[e.g.][]{Gordon13,Bartels16,Lee16}. The 3.5~keV line also has alternative explanations, including known astrophysical lines \citep{Carlson:2014lla,Jeltema15} and charge exchange in sulphur ions  \mbox{\citep{Gu15, Shah16}}. In addition, a recent study using the {\it Hitomi} satellite ruled out the reported detection in Perseus  \mbox{\citep{HitomiC16}}. However, that detection had been anomalously bright compared to the expected mass content of Perseus, and the {\it Hitomi} observation was not long enough to detect the likely flux inferred for Perseus from other clusters.  
 
 We can supplement our knowledge about the viability of these dark matter candidates, and thus the claimed signals, by examining their effects on the properties of cosmic structure formation. In particular, the generation mechanisms for the two candidates mentioned above, namely WIMPs and sterile neutrinos, yield very different momentum distributions for the dark matter. WIMPs are generated thermally, at non-relativistic temperatures~\mbox{\citep{Lee:1977ua}}. Their non-relativistic velocities at decoupling result in WIMPs being gravitationally confined to perturbations that are much smaller than those that form galaxies  \mbox{\citep[e.g.][]{Diemand_05,Schneider13}}. They are thus a `cold' dark matter (CDM) candidate. Sterile neutrinos instead are generated relativistically, due to their much smaller masses, and free-stream out of perturbations of the scale of dwarf galaxies and below  \citep{Dodelson94,Shi99,Abazajian:2001nj,Asaka:06c,Laine08, Ghiglieri15,Venumadhav16} and therefore behave as `warm' dark matter (WDM). Both the free-streaming scale and the radiative decay rate are expressed through the same set of sterile neutrino parameters, linking structure formation directly to the observed X-ray signal. The properties of dwarf galaxies thus provide an opportunity to compare WIMPs and sterile neutrinos in a fashion complementary to indirect searches.
 
 This progress in describing and potentially detecting different dark matter candidates happens to coincide with improvements in our understanding of the region of space in which the properties of the dark matter may be most strongly manifest: the Local Group of galaxies. Since the turn of the century, many new satellite galaxies have been discovered, including  $\sim15$ new Milky Way satellites by the SDSS survey  \mbox{\citep[e.g.][]{Willman05b, Walsh07}}, $\sim11$ Milky Way and Large Magellanic Cloud (LMC) satellites in the DES survey  \citep{Bechtol15,DrlicaWagner15}, and  a series of new M31 satellites in the PAndAS  \mbox{\citep{Martin_09,Richardson11}} survey. Spectroscopic follow up of some of these satellites has revealed that they contain much more dark matter than baryonic matter  \citep{Walker09,Walker10,Wolf10,Tollerud12,Walker16}. More recently, studies have shown that the Large Magellanic Cloud (LMC) satellite of the Milky Way may contain much more dark matter than had previously been thought  \mbox{\citep{Penarrubia16}} and may have brought satellites of its own into the Milky Way system  \citep{Bechtol15,DrlicaWagner15,Koposov15,Jethwa16}.

The effect of sterile neutrinos and WDM has been investigated with dark matter-only simulations and with Monte Carlo tree methods. Studies have shown that the subhalo abundance is strongly dependent on the dark matter properties, and thus limits can be placed on its parameter values \citep{Bode01,Polisensky2011,Abazajian14,Kennedy14,Lovell14,Horiuchi16}. The halo density profiles are also affected, with $\sim$pc-scale cores \citep{Maccio12,Maccio13,Shao13} and $\sim$kpc-sized regions in which the halo densities are suppressed relative to CDM predictions and may prove a better fit to observations \citep{Lovell12,Schneider14,Horiuchi16,Lovell17}. However, most of these studies do not include any input from baryon models, and those that do use results from CDM simulations. It is thus important to check whether the CDM approach is applicable to WDM models, e.g. changes in dark matter profiles due to stripping and feedback, and also to use hydrodynamical models to make predictions for properties not included in dark matter-only simulations, such as satellite ages and metallicities.
 
 We thus require a method to calculate what the properties of the Local Group would be if the dark matter were composed of either WIMPS or sterile neutrinos while taking baryon physics into account. It has been shown that the APOSTLE simulations of the Local Group assuming CDM and a state-of-the-art galaxy formation model provide a good fit to Local Group observations such as luminosity functions and satellite densities  \mbox{\citep{Sawala16}}, and that this model also successfully describes the properties of much larger galaxies \citep{Schaye15}. This methodology is easily extended to WDM candidates such as sterile neutrinos. In this paper we resimulate Local Group-like systems with a selection of sterile neutrino models, and consider under what conditions these models may either be a better fit to the observational data than CDM or such a poor fit as to be ruled out.
 
 This paper is organized as follows. We introduce our simulation suite in Section~\ref{Sims}, present our results in Section~\ref{Res}, and finally draw conclusions in Section~\ref{Con}.

  \section{Simulations}
  \label{Sims}

\subsection{Initial conditions}

We introduce 12 smoothed particle hydrodynamical (SPH) simulations of Local Group-analogue systems that are simulated assuming sterile neutrino dark matter. The initial conditions are modified versions of those used for the first six CDM simulation volumes (which we label AP-1-AP-6) of the APOSTLE project \citep{Fattahi16b, Sawala16}, which were selected from subvolumes of a parent simulation containing pairs of haloes that approximately replicate the masses of the M31 and Milky Way (hereafter abbreviated as MW) dark matter haloes as well as the distance between them and their relative radial and tangential velocities. The maximum softening length for all six runs is $\epsilon=307$~pc; the softening is implemented in proper parsecs at $z<2.8$ and comoving parsecs at $z>2.8$. The initial gas particle mass is $\sim10^5\Msun$; see Table~\ref{TabSim} for the mass in each individual run.

 The cosmological parameters are consistent with the {\em WMAP-7} cosmic microwave background measurements \citep{wmap11}, and take the following values: Hubble parameter $H_0=70.4$~\kms, matter density parameter $\Omega_\rmn{M}=0.272$, baryon density parameter $\Omega_\rmn{b}=0.0455$, dark energy density parameter $\Omega_{\Lambda}=0.728$, linear power spectrum amplitude $\sigma_8=0.81$, and power-law spectral index $n_\rmn{s}=0.967$. It has been suggested that the choice of cosmological parameters has a non-negligible effect on the abundance and structure of satellites in dark matter-only simulations \citep{Springel08b,Polisensky14}. However, the magnitude of these effects is much smaller than the uncertainty in the galaxy formation model parameters. Therefore we do not consider what impact choosing an alternative cosmology, such as \emph{Planck} \citep{PlanckCP13}, would have on our results.

 \begin{table*}
    \centering
    \begin{tabular}{|l|c|c|c|c|c|c|}
      \hline
      Simulation & $m_\rmn{gas}$ ($10^5\Msun$)  &  $M_{\rmn{M31},200}$ ($10^{12}\Msun$)& $M_{\rmn{MW},200}$  ($10^{12}\Msun$) &  $M_{\rmn{M31},*}$ ($10^{10}\Msun$) & $M_{\rmn{MW},*}$  ($10^{10}\Msun$)  & $d_{\rmn{M31-MW}}$ (Mpc) \\
      \hline  
      AP-1-CDM & 1.00 &   1.64 & 1.02 & 4.88 & 1.92 & 0.84 \\
    AP-1-LA10 & 1.00 &   1.62 & 0.98 & 4.72 & 1.83 & 0.82 \\
    AP-1-LA120 & 1.00 &   1.63 & 0.94 & 4.54 & 1.71 & 0.78 \\
    \\
    
      AP-2-CDM & 1.04 &   0.76 & 0.74 & 2.35 & 1.47 & 0.82 \\ 
	AP-2-LA10 & 1.04 &   0.76 & 0.74 & 2.50 & 1.52 & 0.83 \\
	AP-2-LA120 & 1.04 &   0.73 & 0.77 & 2.16 & 1.84 & 0.81 \\
	\\
	
       AP-3-CDM & 1.04 &   1.50 & 1.20 & 2.23 & 2.08 & 0.91 \\
       	AP-3-LA10 & 1.04 &   1.55 & 1.27 & 2.69 & 2.19 & 0.91 \\
       AP-3-LA120 & 1.04 &   1.48 & 1.43 & 2.22 & 1.30 & 0.82 \\
\\	
	AP-4-CDM & 1.02 &   1.26 & 1.25 & 3.60 & 3.20 & 0.84 \\
	AP-4-LA10 & 1.02 &   1.32 & 1.26 & 3.32 & 3.39 & 0.93 \\
	AP-4-LA120 & 1.02 &   1.53 & 1.40 & 3.49 & 3.57 & 0.91 \\
\\		
	AP-5-CDM & 1.04 &   0.89 & 0.84 & 2.39 & 2.21 & 0.82 \\
	AP-5-LA10 & 1.04 &   0.92 & 0.81 & 2.48 & 2.12 & 0.81 \\
	AP-5-LA120 & 1.04 &   0.89 & 0.82 & 2.53 & 1.90 & 0.81 \\
\\
	AP-6-CDM & 1.06 &   2.15 & 1.14 & 4.48 & 2.32 & 0.94 \\
	AP-6-LA10 & 1.06 &   2.01 & 1.03 & 3.49 & 2.53 & 0.95 \\
	AP-6-LA120 & 1.06 &   1.90 & 1.03 & 3.01 & 2.12 & 0.95 \\

      \hline
      Local Group & n/a & $ -- $& $ -- $ &  $(10-15)^{a}$& $6.08\pm1.14^{b}$ &$0.785\pm0.025^{c}$ \\
     
      \hline
    \end{tabular}
    \caption{Important parameters of each simulation: simulation name, initial gas particle mass, $m_\rmn{gas}$, virial masses of the two central haloes, $M_{\rmn{M31},200}$ and $M_{\rmn{MW},200}$, 30~kpc aperture stellar masses of these haloes, $M_{\rmn{M31},*}$ and $M_{\rmn{MW},*}$ and the halo separation $d_{\rmn{M31-MW}}$. We also include the measured values for the Local Group, which are obtained from: $^{a}$\citet{Tamm12}, $^{b}$\citet{Licquia15},$^c$\citet{McConnachie05}. The discrepancy between the simulated and observed stellar masses is discussed in the main text. We do not include estimates for the halo masses because the uncertainties on these measurements are very large. A more comprehensive presentation of the properties of each of the CDM simulations can be found in \citet{Fattahi16b}.}
    \label{TabSim}
  \end{table*}

The modification applied to the APOSTLE initial conditions for this study is that the wave amplitudes are rescaled to a sterile neutrino dark matter power spectrum. The properties of this power spectrum are set by two particle physics properties: the sterile neutrino mass and the lepton asymmetry with which the sterile neutrinos are generated \citep{Shi99,Laine08,Venumadhav16}. We denote the lepton asymmetry using the parameter $L_6$, which is defined as $L_6\equiv 10^6(n_{\nu_e}-\bar n_{\nu_e})/s$, where $n_{\nu_e}$ is the lepton number density, $\bar n_{\nu_e}$ the anti-lepton number density, and $s$ the entropy density. This parameter space is bounded on all sides due to considerations of X-ray non-detections, structure formation and dark matter abundance \citep{Boyarsky09a,Lovell16,Schneider16}. However, attempting to scan the allowed range is prohibitive for high-resolution hydrodynamical simulations. We therefore motivate our choice from the interpretation of the detected 3.55~keV line as the decay of a 7.1~keV sterile neutrino. 

 If a sterile neutrino of this mass constitutes the entirety of the dark matter, the allowed range of \Ls can be inferred from the X-ray observations and fluxes of  \citet{Boyarsky14a,Boyarsky15}, \citet{Bulbul14}, \citet{Iakubovskyi15} and \citet{Ruchayskiy16} via estimates of the mixing angle that regulates the decay rate, denoted \Ss. The amplitude of the flux, coupled to the projected mass density of the target and the line position, gives \Ss$=[2.1,8.1]\times10^{-11}$. Numerical calculations for the production of sterile neutrinos performed using the codes of \citet{Laine08} and \citet{Ghiglieri15} \citep[see][for an alternative calculation]{Abazajian14,Venumadhav16} find that this mixing angle for a 7~keV sterile neutrino corresponds to $L_6=[9,11.2]$.

In this paper we use a 7.0~keV sterile neutrino with two values of the lepton asymmetry: \Ls=10 (\Ss=$3.7\times10^{-11}$), which marks the centre of the 3.5~keV line expected region, and \Ls=120 (\Ss=$8.0\times10^{-13}$), which is close to the spectrum with the lowest wavenumber cutoff attainable for any 7~keV sterile neutrino, although this mixing angle is too small to be responsible for the measured flux of the 3.5~keV line. We expect that the difference between 7.0 and 7.1~keV is negligible for our results.

\begin{figure}
    \includegraphics[scale=0.33]{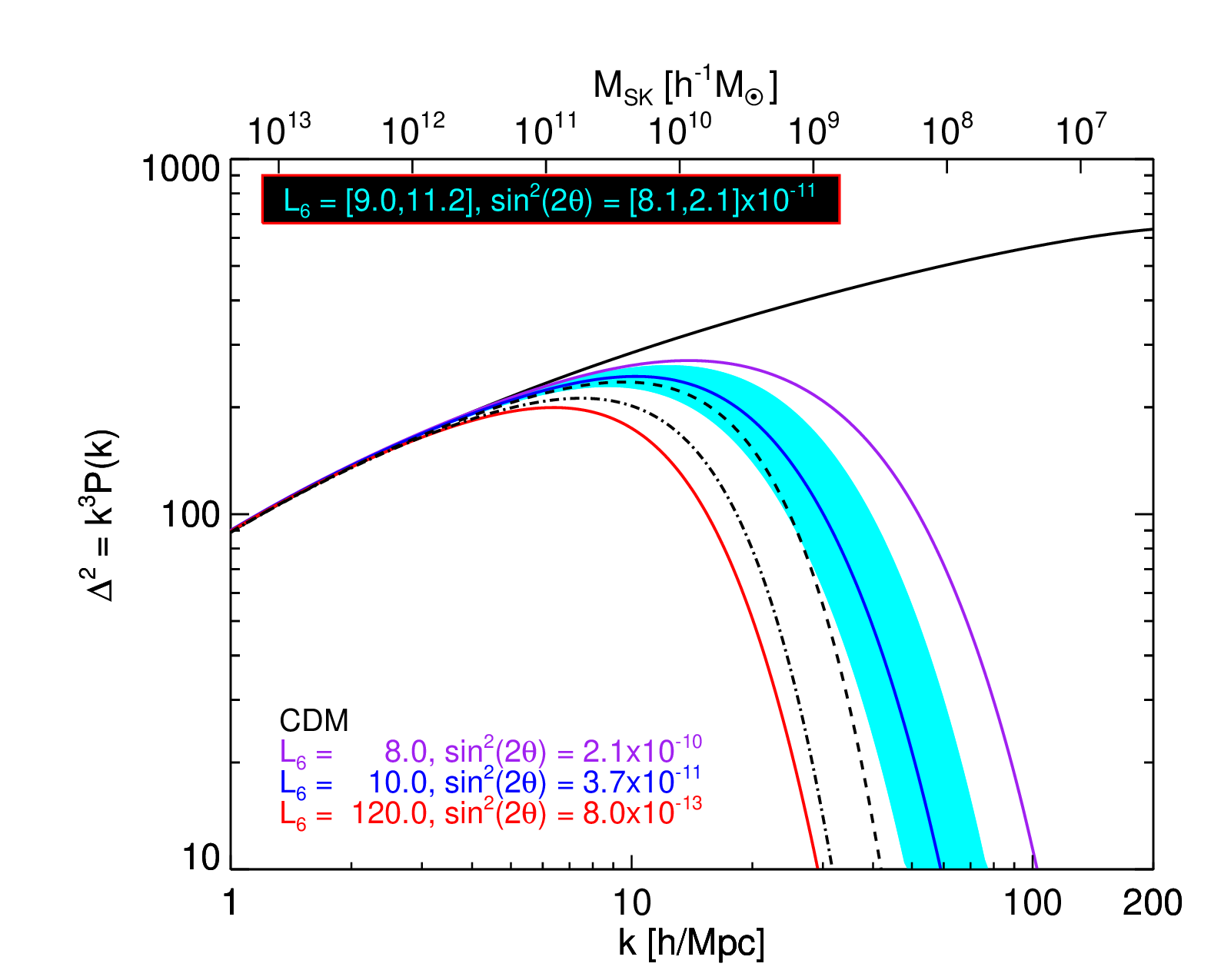}
    \caption{The $z=0$ linear theory matter power spectrum for a series of dark
    matter models. The solid black line represents CDM. Three 7~keV sterile
    neutrino models are shown with different values of the lepton
    asymmetry. The purple, blue and red lines denote lepton
    asymmetry $L_{6}$ = 8, 10 and 120 respectively. The cyan region marks the limits for the family of sterile neutrino models that are consistent with the 3.5~keV line, i.e. between $L_6=9$ and $L_6=11.2$. We indicate the mixing angle, $\sin^2(2\theta)$, that corresponds to each of the models in the legend. For reference we include the power spectrum for a 2~keV thermal relic as the dot--dashed, black line and a 2.5keV thermal relic as the dashed line. }
    \label{PowerSpectra}
  \end{figure}

 We plot the dimensionless linear theory matter power spectra for these values of $L_6$, along with four others, in Fig.~\ref{PowerSpectra}. We also plot CDM, the 7~keV spectra whose amplitude peaks at the highest wavenumber of any 7~keV sterile neutrino (\Ls=8.0),  and those that mark the upper (\Ls=9.0) and lower (\Ls=11.2) ranges expected from the published detections and exclusion limits from M31, Perseus, the Galactic Centre and the Draco dSph. On the upper $x$-axis we show the mass scale associated with these wavenumbers, as calculated using a sharp $k$-space filter \citep{Benson13}. All of our models bar $L_6=120$ have more small scale power than the `fiducial' 2~keV thermal relic used in most WDM studies. Those with $8<L_6\lsim11$ also have more power than the \citet{Viel_13} $3\sigma$ limit on the thermal relic mass (2.5~keV) and are therefore not in strong tension with the Lyman-$\alpha$ observations \citep[but see also][]{Irsic17}.
 
 \subsection{Simulation code and halo identification}
 
 In all respects apart from the choice of matter power spectra, these simulations are performed in the same way as the original APOSTLE runs. The code used is the modified version of \textsc{p-gadget3} developed for the {\sc eagle} project \citep{Crain15,Schaye15}. This setup features the {\sc anarchy} implementation of SPH (Dalla Vecchia~et~al., in prep., see also \citealp{Schaller15}), radiative cooling in the presence of a UV ionizing background that is spatially uniform and evolves with redshift \citep{Haardt01, Wiersma09a}, black hole growth \citep{Springel05b,RosasGuevara15}, star formation \mbox{\citep{Schaye08}} and the feedback associated with the formation of stars and black hole growth  \mbox{\citep{Booth09, DallaVecchia12}}.  The effects of these processes often cannot be calculated from first principles, and therefore must be modelled using so-called `subgrid physics'. In particular, the values of the parameters that control the stellar and AGN feedback must be fixed by calibration against a set of observations, and may have to be re-calibrated if runs are performed at other resolutions. Predictive power of the model then derives from comparison against other observables. In the {\sc eagle} project, the subgrid model parameter controlling the feedback efficiency has been calibrated to reproduce three observables: the $z=0.1$ galaxy stellar mass function, the sizes of disc galaxies, and the amplitude of the black hole-stellar mass relation. 
 
  The choice of parameters that describe the galaxy formation model is that of the {\sc eagle} L0100-N1504 run (otherwise known as the REFERENCE model). These parameters were calibrated exclusively using the CDM matter power spectrum: a proper application of the {\sc eagle} calibration philosophy dictates that we should recalibrate our model to find the best-fitting model parameters for our sterile neutrino power spectra. However, the observations upon which the subgrid parameters are calibrated are on scales where there is little difference between CDM and WDM models (stellar mass $>10^{8}\Msun$) and so we can safely use the original values. In addition, this approach allows us to make predictions for the satellite luminosity functions. This would not be possible if we had calibrated to these data for each dark matter model.

 We identify haloes and subhaloes using the {\sc subfind} algorithm \citep{Springel01,Dolag09}. Following \citet{Schaye15}, galaxy stellar masses are determined by calculating the total stellar mass enclosed within a 30~kpc spherical aperture about the (sub)halo centre, which we denote hereafter $M_{*}$. In general, the properties of a galaxy are considered to be resolved within its host halo if  $M_{*}>10^{7}\Msun$, which corresponds to approximately 100 star particles (but see \citealp{Schaye15} for the different numbers of particles required to resolve different properties.) For lower stellar masses, the properties of individual satellites are poorly resolved. However, the total number of satellites with a given stellar mass is well resolved \citep{Sawala16}, and the existence of a galaxy within a halo is well established by the presence of even just one star particle, due to the requirement of many gas particles to form a star particle. We thus select different stellar mass limits for our various measurements. Where we consider only the total stellar mass of the satellite, we accept all satellites with $M_{*}>10^5\Msun$ ($\sim1$ particle). In our analysis of the galaxies' metallicities and stellar ages, we examine satellites of $M_{*}>10^6\Msun$ ($\sim10$ particles). Occasionally the subhalo finder will misidentify fragments of one galaxy as a separate galaxy in its own right. We follow the lead of \citet{Schaye15} in that we merge together galaxies whose separation is less than the minimum of the stellar half-mass radius and 3~kpc. Note that WDM runs also suffer from the formation of spurious haloes; we discuss this issue in Appendix~\ref{app:ss}.  
 
 In Table~\ref{TabSim} we present some important quantities for each of our runs: the simulation names, the M31 and MW halo virial masses -- calculated as the total mass within the radius enclosing a mean density 200 times the critical density --along with their 30~kpc aperture stellar masses, and the halo separations. In the CDM case, we define the M31 halo as that with the larger virial mass or, in those cases where the two galaxies reside in the same Friends-of-Friends halo, the galaxy in the more massive subhalo. Our simulation names are built using the volume identifier (AP-1-AP-6) and the dark matter model (LA10 as a shorthand for  7~keV, \Ls=10 sterile neutrino, LA120 for 7~keV, \Ls=120 and the original CDM). We also include the observed values for the MW and M31. As shown by \citet{Sawala16}, the simulated halo masses are consistent with the observations, but the stellar masses for both galaxies are lower than those inferred for the MW and M31 by at least a factor of 2. This is due to overly strong feedback in $\sim10^{12}\Msun$ mass haloes in the {\sc eagle} model \citep{Schaye15}. 
 
 Using these quantities we can assess to what degree our sterile neutrino runs still host systems that look like the Local Group, given that alterations in the small scale power can change the formation histories of the large haloes for sufficiently warm models \citep{Libeskind13}. The change in halo separation between the CDM, LA10 and LA120 versions or `realizations', of the same volume is never greater than 11 per cent, and typically less than 1 per cent. The halo virial mass changes by a maximum of 6.5 per cent and often much less. The stellar mass changes by a larger amount, and is typically lower in the LA10/LA120 galaxies; however, the greatest variation is 22 per cent and the median difference between the CDM and LA10/LA120 stellar masses is only 5.6 per cent. Therefore, we are confident that the large scale, Local Group-like properties of these systems are preserved in the LA10 and LA120 runs. Note that in the rest of this work we treat the MW and M31 analogues as equally likely to host MW-like systems. 

  \section{Results}
  \label{Res}
  
    \subsection{Satellite structure}
  \label{ss:struc}
  \subsubsection{Satellite masses}
  \label{ss:smass}
  
  We begin our discussion of the results with the relationship between the stellar mass and total/dark matter mass for isolated galaxies and satellite galaxies. We define a satellite as a galaxy that is located within 300~kpc of a MW or M31 analogue. 
  
\begin{figure*}
    \includegraphics[scale=0.55]{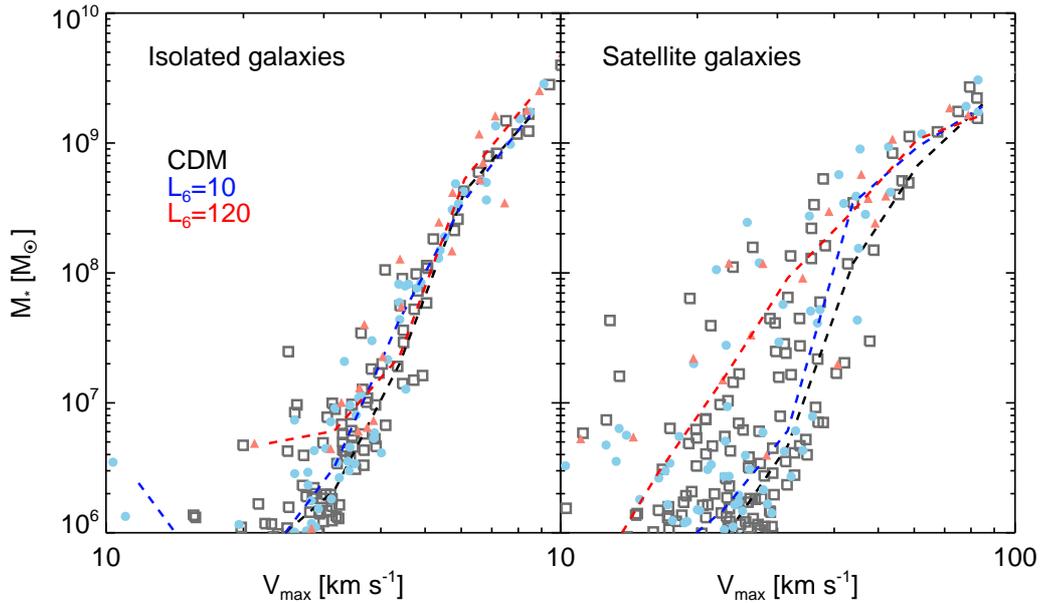}
    \caption{Galaxy stellar mass as a function of \vmax for isolated galaxies (left-hand panel) and satellites (right-hand panel). The black squares, blue circles and red triangles denote the CDM, LA10 and LA120 systems respectively. We also show the median trends for the three models. The shade of the data points has been faded in order to make the median trends visible.}
    \label{MsubMstar}
  \end{figure*}

  The evolution of satellite galaxy structure depends on the properties of the haloes in which they form and on how they are processed after accretion on to the parent halo. WDM subhaloes lose a greater proportion of their mass through stripping than CDM subhaloes do, due to their lower central densities \citep{Lovell12}. Differences between isolated galaxies -- which can be thought of as analogues for satellites pre-infall -- have been investigated by \citet{Bose16a}. In Fig.~\ref{MsubMstar} we plot the galaxy stellar mass as a function of \vmax, which is defined as the peak of the halo circular velocity curve; the galaxies are separated into isolated galaxies and satellites. We use this quantity as a proxy for the halo mass, since the mass itself is poorly defined for subhaloes and difficult to measure for satellites close to the host halo centre. We define isolated galaxies as those that are located within 2~Mpc of the MW-M31 barycentre but more than 300~kpc from both the MW and M31 analogues. 
  
The isolated galaxies in all three models exhibit a steep correlation between stellar mass  with \vmax. The scatter in \vmax at fixed stellar mass is much better constrained than the scatter in stellar mass at fixed \vmax.  There is a very slight preference in the data for the warmer model haloes to have larger stellar masses than their CDM counterparts for \vmax $<50$~\kms. Some satellite galaxies have stellar masses over an order of magnitude larger than isolated galaxies at the same \vmax, due to the stripping of the outer layers of the dark matter halo \citep{Sawala16b} whilst the centrally concentrated, stellar component remains intact. The relative increase in stellar mass per unit total mass is stronger still in the WDM satellites, in agreement with our expectation that lower dark matter densities facilitate a higher rate of stripping. 
    
\begin{figure}
    \includegraphics[scale=0.33]{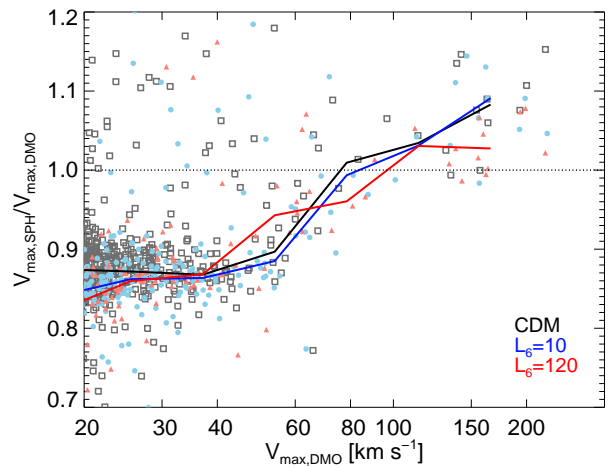}
   \caption{Ratio of \vmax between the SPH and the matched dark matter-only versions of our  isolated haloes, as a function of the dark matter-only \vmax. CDM, LA10 and LA120 haloes are shown as black squares, blue circles,and red triangles respectively. The corresponding solid lines denote the median relations. We restrict our halo pair selection to those in which the SPH counterpart has \vmax$>20~$\kms, is no further than 2~Mpc from the MW-M31 barycentre, and the largest constituent subhalo has a bound mass $>0.8$~$M_{200}$ in order to exclude merging systems.}
     \label{VmaxComp}
  \end{figure}

    Evidence for the effect of baryon physics on these haloes can be obtained by comparing the results of the SPH simulations with $N$-body realisations of the same volume in which all of the matter is assumed to behave as a collisionless fluid, known (somewhat inaccurately) as a `dark matter-only' run. \citet{Sawala16} matched pairs of haloes between dark matter-only and hydrodynamical runs of the same simulation volume, and showed that haloes with a maximum circular velocity, \vmax$<30$~\kms in the dark matter-only run have their values of \vmax suppressed by  12 per cent in the median when the baryons are included. They ascribe this effect to the loss of gas due to heating by reionization and stellar feedback; the lower total mass leads to slower growth of the halo. We repeat this exercise for the isolated haloes in our runs, making use of dark matter-only versions of our SPH simulations, and plot the results in Fig.~\ref{VmaxComp}. Remarkably, there is very little difference between the sterile neutrino and CDM models, in that the same degree of \vmax suppression is found in both. In all three models we detect an increase in \vmax for the hydrodynamical runs at \vmax$>100$~\kms, which is due to dissipational collapse of the baryons and its effect on the dark matter \citep{Sawala13,Schaller15c}.

 \subsubsection{Satellite concentrations}
  \label{ss:sconc}

\begin{figure*}
    \includegraphics[scale=0.50]{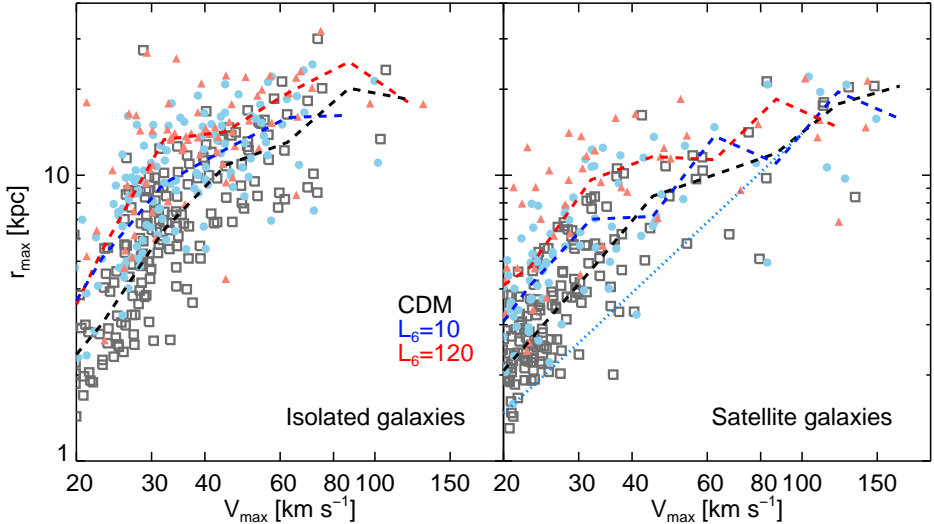}
    \caption{Scatter plot of $V_\rmn{max}$ and $r_\rmn{max}$ for isolated galaxies (left-hand panel) and satellites (right-hand panel). Black squares denote CDM, blue circles LA10, and red triangles LA120. The median relations are shown as dashed lines. The dotted grey line in the right hand panel shows the mean relations for CDM dark matter-only satellites as derived by \citet{Springel08b} in the {\emph WMAP}1 cosmology.}
    \label{VmRm}
  \end{figure*}

Once the effects of baryon physics have been taken into account, the mass evolution of subhaloes is dominated by stripping, the rate of which is linked to the concentration of the halo. Even without stripping, haloes with lower concentrations may be in better agreement with measurements of satellite velocity dispersions, and avoid the so-called too big to fail problem in CDM (\citealp{BoylanKolchin11,BoylanKolchin12} but see \citealp{Fattahi16}). This concentration can be assessed by comparing $V_\rmn{max}$ with the radius at which this velocity occurs, $r_\rmn{max}$. In Fig.~\ref{VmRm} we plot these quantities for our isolated and satellite galaxies. In the isolated galaxy case, we find that there is a clear trend for LA120 haloes to have the largest $r_\rmn{max}$ at a given $V_\rmn{max}$, and thus lower concentrations, which is consistent with expectations from dark matter-only simulations. This trend between the dark matter models is continued in the panel showing the satellite galaxies. All three dark matter models show lower concentrations than the mean relation derived for satellites in a dark matter-only simulation by \citet{Springel08b}. An offset in the median relation for LA120 and LA10 relative to CDM is still evident for $V_\rmn{max}<40~\rmn{km~s}^{-1}$. For both classes of galaxies the overlap between the two distributions is considerable despite the different stripping rates; we defer analysis of the stripping rates in satellite merger trees to a future paper.

  \subsubsection{Satellite central densities}
  \label{ss:cendens}

In practice, measurements of the central densities of MW and M31 satellites are restricted to the half-light radii, denoted $r_{1/2}$, the values of which are $\leq1$~kpc \citep{Walker09, Wolf10} and therefore it is not possible to measure $r_\rmn{max}$ and $V_\rmn{max}$ directly. Instead, inferences about these quantities are often made using high resolution dark matter-only \citep{BoylanKolchin12,Lovell12,Horiuchi16} or, more rarely, given the high computational costs, hydrodynamical simulations with sub-kiloparsec resolution \citep{Brooks2014,Governato15,Fattahi16,Sawala16} to estimate the enclosed mass within the satellite's half-light radii. For the brightest MW satellites, the half-light radii range from 300~pc to 1~kpc, length scales that are below the softening scale of our simulations. Therefore, we here consider the enclosed masses at the scale of 1~kpc, which we parametrize as the value of the circular velocity curve at 1~kpc, $V_{1~\rmn{kpc}}$. In order to check for the convergence of this quantity, we matched isolated haloes in the AP-1-CDM run with those in a higher resolution simulation of the same volume ($\times8$ more high resolution particles, presented as resolution L1 in \citealp{Sawala16}). We find that these masses are suppressed in AP-1-CDM runs relative to their higher resolution counterparts, by up to 20~per~cent for haloes with  $V_{1~\rmn{kpc}}>20$~\kms and as much as 40~per~cent for haloes with $V_{\sim1~\rmn{kpc}}>10$~\kms.
 
\begin{figure}
   \includegraphics[scale=0.33]{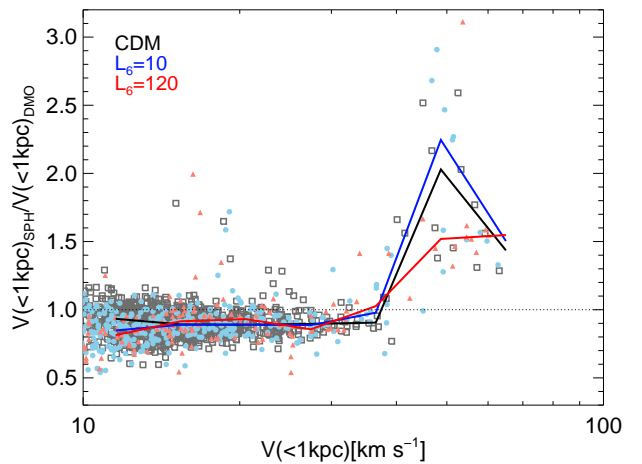}
   \caption{Ratio of mass enclosed within a radius of 1~kpc as a function of the matched dark matter-only-hydrodynamic pairs as introduced in Fig.~\ref{VmaxComp}. }
     \label{M2DMO}
  \end{figure}

 Our expectation is that these central densities will be lower for our LA10 and LA120 models due to the later formation times observed in dark matter-only simulations. As shown in Fig.~\ref{VmaxComp} \vmax was uniformly suppressed by 12 per cent for small haloes irrespective of the dark matter model chosen. However, it is conceivable that the mass profile in the inner parts of the halo may respond differently, depending on the choice of dark matter. We therefore calculate the suppression in the circular velocity profile evaluated at 1~kpc, $V_{1~\rmn{kpc}}$, due to baryon effects using our dark matter-only-SPH halo pairs, and show the results in Fig.~\ref{M2DMO}. We find that all three models show the same suppression in central mass below $V_{1~\rmn{kpc}}<30$~\kms, namely $\approx10$~per cent. The scatter is of order 10~per~cent, although not larger when considering the entire subhalo. We therefore conclude that the baryon physics model has the same impact on our sterile neutrino models as it does on CDM. 
 
\begin{figure*}
    \includegraphics[scale=0.5]{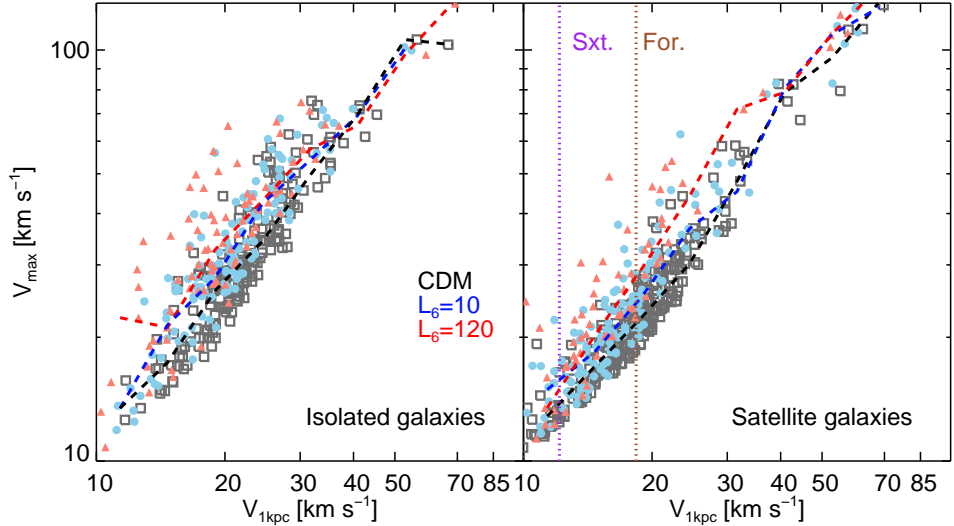}
    \caption{Galaxy \vmax as a function of the circular velocity at 1~kpc for isolated galaxies (left-hand panel) and satellites (right-hand panel). The scheme for plotting the data points and median trends is the same as in Fig.~\ref{VmaxComp}. The brown and purple dotted lines show the measured values of $V_{c}(r_{1/2})$ for the Fornax and Sextans satellites respectively.}
    \label{V1Vm}
  \end{figure*}

We end this subsection by making a connection to observations. As stated above, the length scales at which satellite densities are measured are up to three times smaller than the smallest scales resolved by our simulation. A competitive analysis would need to either resolve these small scales \citep{BoylanKolchin12,Lovell12,Fattahi16,Sawala16} or employ extrapolations to small scales along with a large sample of host haloes \citep{JiangF15,Lovell17}. We restrict ourselves to the relationship between \vmax and $V_\rmn{1kpc}$ for bright ($>10^{5}\Msun$) satellites. In Fig.~\ref{V1Vm} we plot the \vmax of isolated and satellite galaxies as a function of $V_\rmn{1kpc}$ for the three models. There is a clear difference between CDM and the sterile neutrino models for $V_\rmn{1kpc}<40$~\kms. The median LA120 isolated \vmax is 40~per~cent higher than that of the median CDM halo with the same $V_\rmn{1kpc}$, and LA10 takes \vmax 20~per~cent higher than CDM.  This pattern persists for the satellites, even though the LA120 satellites are more heavily stripped. 

The ability of each model to match the observed satellite galaxy densities is shown by plotting the circular velocities for two satellites whose measured half-light radii are approximately 1~kpc: Fornax ($r_{1/2}=944$~pc) and Sextans ($r_{1/2}=1019$~pc) \citep{Irwin95,Wolf10}. The derived values of $V_{1/2}\approx V_\rmn{1kpc}$ are 18 and 12~\kms respectively \citet{Walker09,Walker10,Wolf10}. On the naive assumption that all satellites of the measured Fornax  $V_\rmn{1kpc}$ are equally likely to host Fornax, regardless of their stellar mass, the most likely \vmax of the host subhalo is several \kms larger in LA10 subhaloes than in CDM subhaloes, and larger still in LA120. There are hints of similar behaviour for potential Sextans hosts, however these lower mass subhaloes are more poorly resolved and so it is not possible to draw any conclusions about which \vmax fits Sextans best for each model. What we can conclude is that sterile neutrino models (and even WDM models more generally) that feature baryons can impact on the agreement between the densities of observed and simulated satellites through both the suppression of their \vmax functions and the lower concentrations of subhaloes with the same \vmax as in CDM. However, higher resolution simulations will be required to determine whether the agreement for WDM models is as good as, or even better than, that shown for CDM. 
 
\begin{figure}
    \includegraphics[scale=0.33]{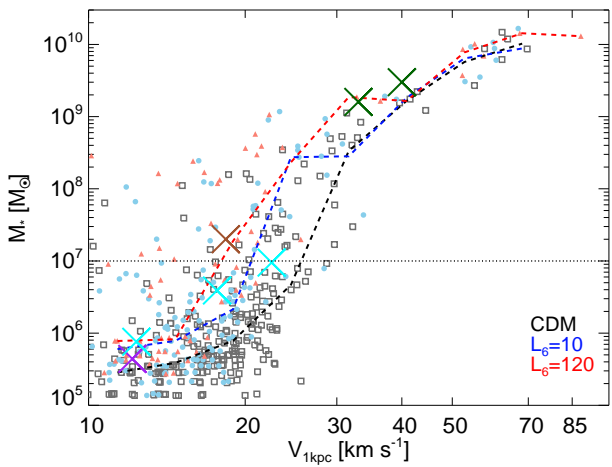}
    \caption{Satellite stellar mass as a function of the circular velocity at 1~kpc. The scheme for plotting the data points and median trends is the same as in Fig.~\ref{VmaxComp}. The brown and purple crosses show the measured values for the Fornax and Sextans satellites respectively. We also include estimates for the LMC and SMC as green crosses, and three Andromeda satellites with half light radii of approximately 1~kpc (AndI, AndVII and AndXXI) as cyan crosses. The dotted line marks the stellar mass above which the satellite properties are well resolved.}
    \label{V1Mstar}
  \end{figure}

  In practice, some subhaloes are more likely to host Fornax-like satellites than others. The stellar mass of Fornax is relatively high, and it is therefore unlikely to be hosted in the smaller \vmax subhaloes, as shown in Fig~\ref{MsubMstar}. In Fig.~\ref{V1Mstar} we plot the stellar mass as a function of $V_\rmn{1kpc}$ for satellite galaxies. There is a clear trend for LA120 to have the brightest satellites at fixed $V_\rmn{1kpc}$ and for CDM to have faintest. All three models produce a population of satellites with similar stellar masses and $V_\rmn{1kpc}$ values to that of Fornax, indeed it has been shown in \citet{Fattahi16} that the CDM APOSTLE simulations predict satellites with the densities and stellar masses of the observed satellites. However, in the CDM model the best matching points are clearly outliers, with most satellites of the same central density over an order of magnitude fainter. Many of these faint satellites are poorly resolved, however we expect that increasing the resolution would result in satellites that have higher $V_\rmn{1kpc}$ (see our discussion of resolution in the CDM L1 simulation above) and are fainter \citep{Schaye15}, therefore this conclusion is robust.
  
   The CDM model predicts many faint satellites with a much lower stellar mass than Fornax yet have the same central density. By contrast, the LA120 model does not generate anywhere near as many faint satellites {at this central density} and therefore Fornax is located at roughly the centre of the simulated distribution: of the 28 LA120 satellites of $V_\rmn{1~kpc}$ within 20~per~cent of that of Fornax, 43~per~cent are brighter than Fornax. LA10 constitutes an intermediate case where Fornax is 10 times brighter than the median satellite of the same $V_\rmn{1kpc}$. 13 per cent of LA10 satellites are brighter than Fornax (out of 68 satellites) compared to 5 per cent of the 144 CDM satellites. We also include points from five other satellites, the LMC (stellar mass estimation of  \mbox{\citealp{vdMarel02}} and $V_\rmn{1kpc}$ estimate from \citealp{Buckley15}) and SMC (stellar mass and $V_\rmn{1kpc}$: \citealp{Bekki09}) and three M31 satellites: AndI, AndVII, and And XXI (stellar mass and $V_\rmn{1/2}$: \citealp{Tollerud12}). These satellites also trace a line that best matches the LA10 result, although less extreme than Fornax. All three models also produce Sextans-like systems; however, they are resolved with fewer than 10 star particles, precluding robust conclusions.   
  
  \subsection{Satellite abundances}
  
\begin{figure*}
    \includegraphics[scale=0.3]{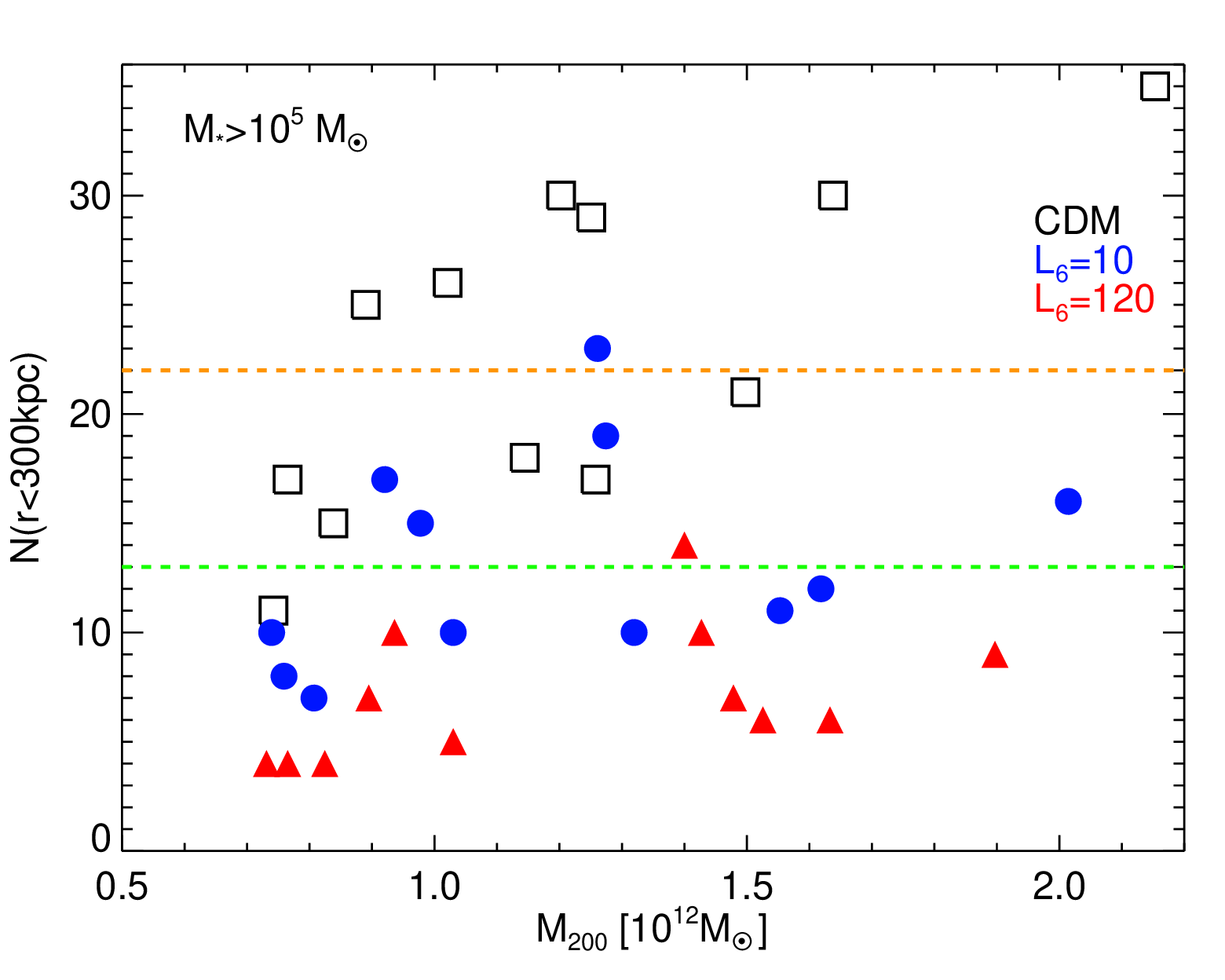}
      \includegraphics[scale=0.3]{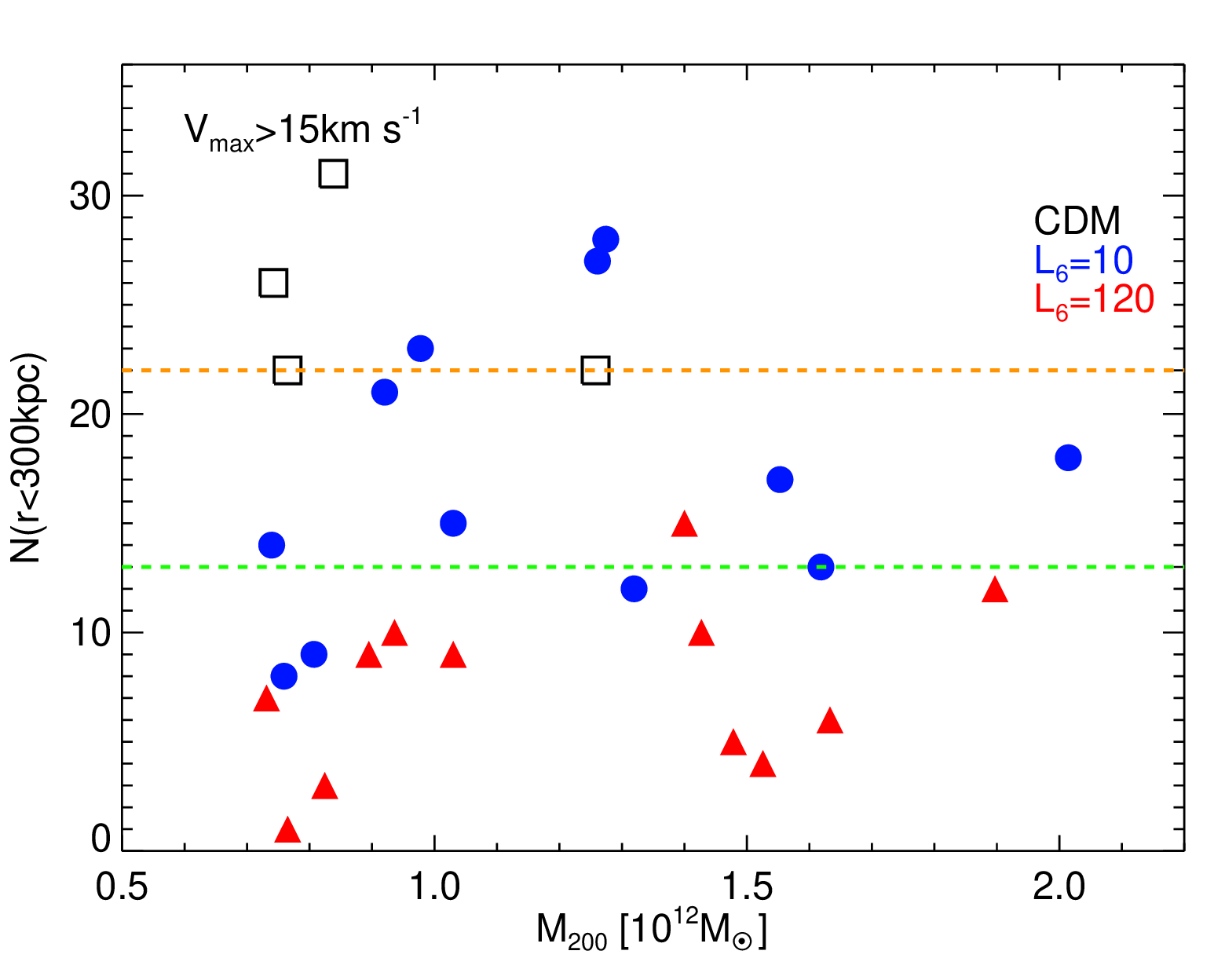}
    \caption{The abundance of galaxies for the satellite systems as a function of $M_{200}$. We define the Local Group mass as the sum of the MW/M31 $M_{200}$ values. The satellites tally includes all satellite galaxies of $M_{*}>10^{5}\Msun$ within 300~kpc. The left-hand panel uses satellites with $M_{*}>10^{5}\Msun$, and the right-hand panel subhaloes with $V_\rmn{max}>15$~\kms, irrespective of whether the subhalo hosts a galaxy. The black squares, blue circles and red triangles denote the CDM, LA10 and LA120 systems respectively. The dashed green lines show the number of observed satellites within the stellar mass and radius limits as compiled by \citet{McConnachie12}; the dashed orange lines show the same quantity for the M31 satellites. We assume that the census of these bright MW satellites, as compiled by \citet{McConnachie12}, is complete.}
    \label{SatLum}
  \end{figure*}

The ability to obtain the correct number of Local Group galaxies in general, and MW/M31 satellites in particular, offers one of the simplest and most important tests for LA10 and LA120. The abundance of these galaxies will also be affected by the total mass of the host dark matter distribution, which is observationally highly uncertain. Therefore,  in Fig.~\ref{SatLum} we plot the number of satellites with stellar masses $>10^{5}\Msun$ for the MW and M31 systems. We use two filters for selecting satellites: stellar mass (left-hand panel of Fig.~\ref{SatLum}) and also \vmax, irrespective of whether the subhalo hosts any star particles (the right-hand panel of the same figure); the second selection enables us to check whether weaker reionization feedback may result in more low mass galaxies. 
  
  In all three models there is a slight trend for more massive central haloes to host more satellites, although the scatter is very large \citep{Fattahi16b}. It is readily apparent that the number of satellites in our sterile neutrino models is suppressed relative to CDM, with a median of six satellites with $>10^5\Msun$ for LA120 and 11 for LA10 compared with 22 for CDM. Recall that \citet{Schaye15} showed that galaxies with fewer than 100 particles -- in this case a stellar mass of $\sim10^{7}\Msun$ -- have poorly sampled supernova rates, thus the number of supernova explosions is lower, the galaxies brighter, and therefore the galaxy number densities are likely overestimated. This does mean, however, that the constraints we place on the sterile neutrino models are likely to be very conservative.
  
   Only one of the LA120 systems produced more satellites than the current number observed, compare to five of the LA10 systems. However, the scatter introduces considerable overlap between the three distributions: there is one LA120 system that produces more satellites than one of the CDM systems, and one LA10 system with a higher satellite abundance than half of CDM. Selecting haloes instead by \vmax makes little difference to the LA120 result, whereas the other two models show a strong increase. 
   
  The M31 galaxy requires more stringent limits, in that only one of the LA10 systems achieves more than 22 satellites without relaxing the feedback requirements. The LA120 model does not produce this many satellites under any circumstances, and therefore we are confident that the M31 halo must have a mass$>2.2\times10^{12}\Msun$ in order for LA120 to be a viable model.
  
\begin{figure*}
     \includegraphics[scale=0.3]{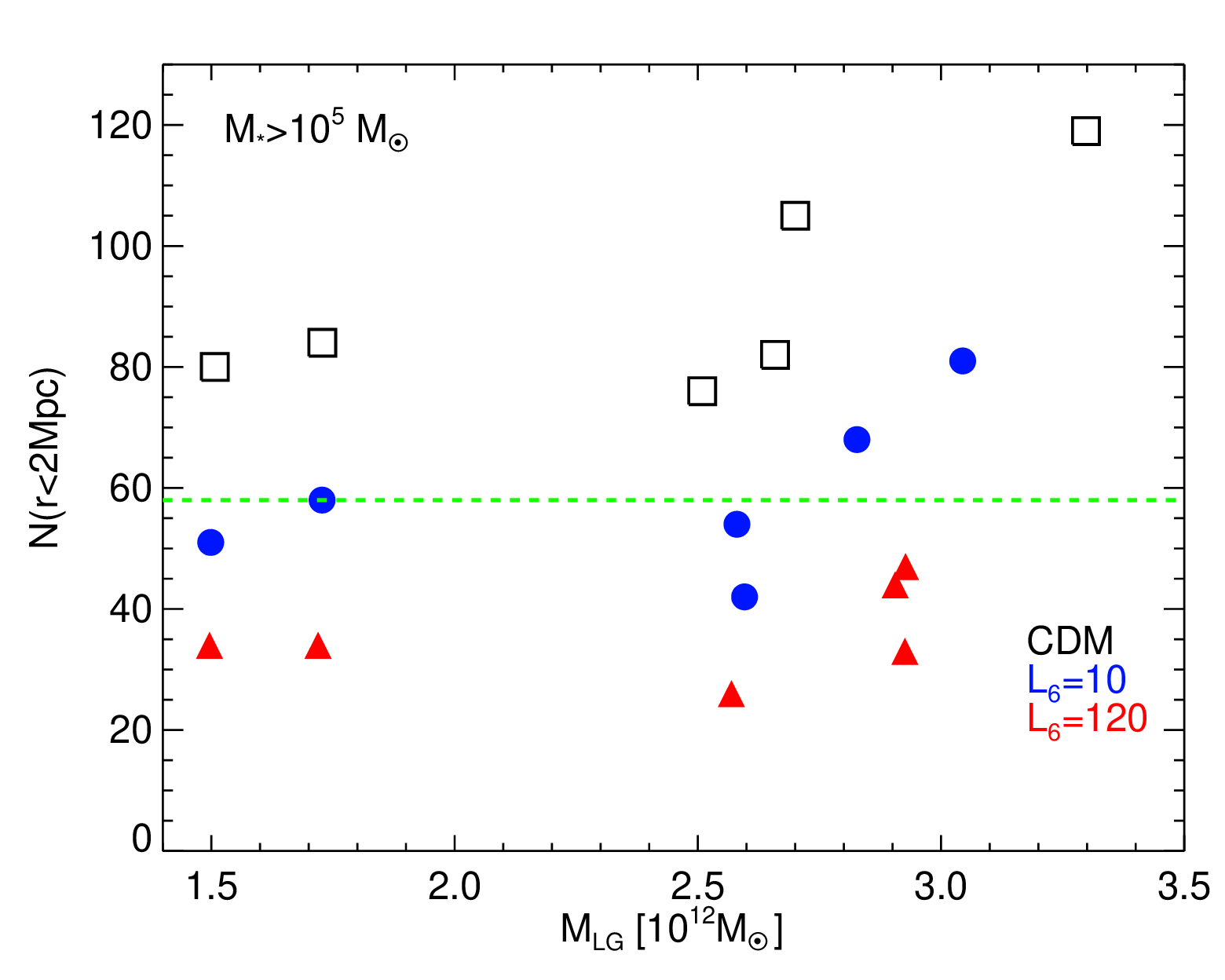}
       \includegraphics[scale=0.3]{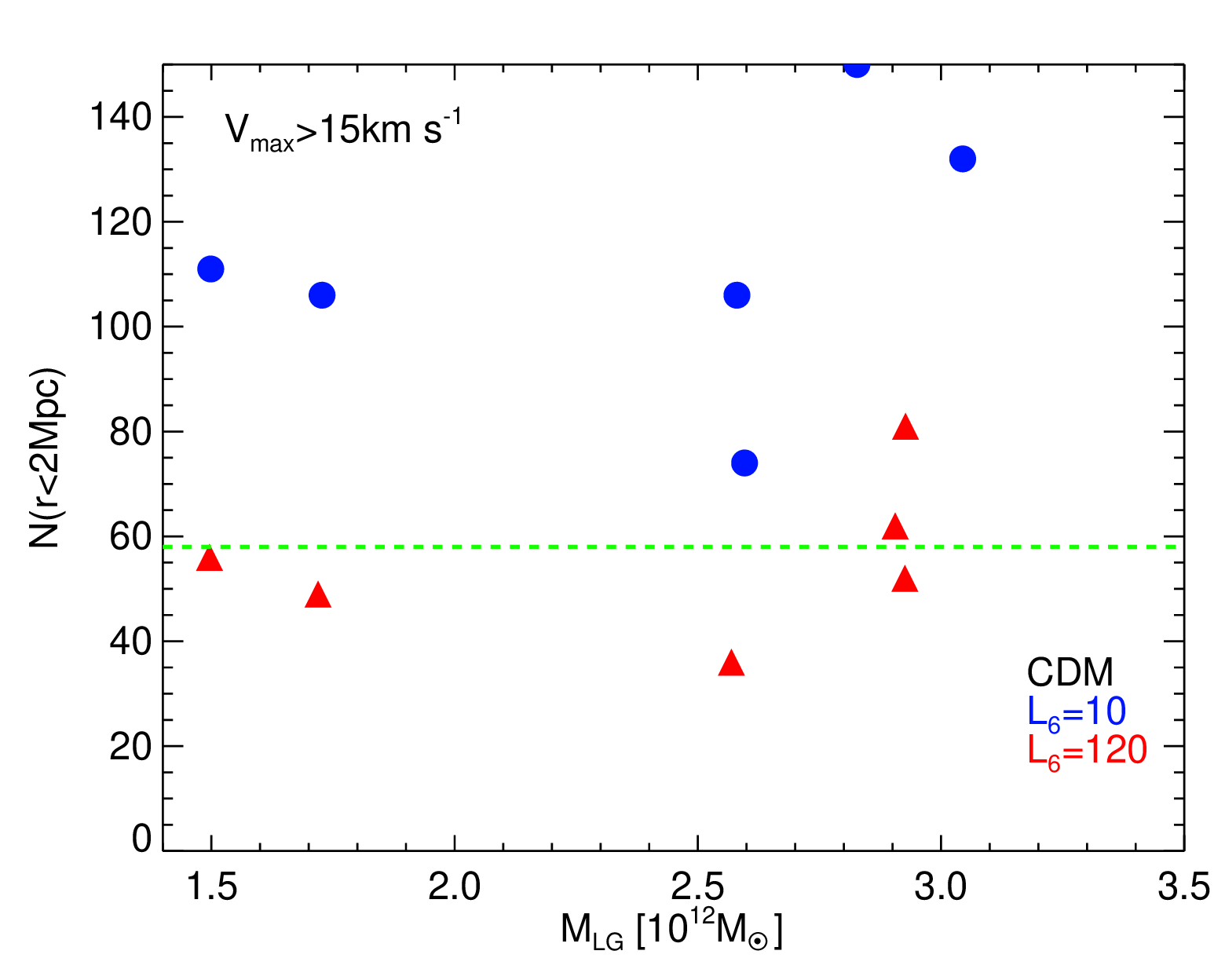}
    \caption{The abundance of galaxies for the Local Groups as a function of the Local Group mass. We define the Local Group mass as the sum of the MW/M31 $M_{200}$ values. The Local Group galaxy count features all galaxies within 2~Mpc of the M31-MW barycentre, including the MW and M31 satellites. The left-hand panel uses satellites with $M_{*}>10^{5}\Msun$, and the right-hand panel subhaloes with $V_\rmn{max}>15$~\kms, irrespective of whether the subhalo hosts a galaxy.  The black squares, blue circles and red triangles denote the CDM, LA10 and LA120 systems respectively. The dashed green lines show the number of observed galaxies within 2~Mpc of the Local Group barycentre within the stellar mass and radius limits as compiled by \citet{McConnachie12}. We do not correct for incompleteness, and therefore these measurements are lower limits on the complete galaxy abundance.}
    \label{LGLum}
  \end{figure*}

In order to decrease the relative shot noise we also consider the total abundance of galaxies within 2~Mpc of the M31-MW barycentre, including satellites, and plot the results in Fig.~\ref{LGLum}. This approach narrows the overlap between the models considerably. The CDM model consistently produces more galaxies than observed, a discrepancy that can be resolved if more satellites of this stellar mass are discovered, or if baryonic feedback in these relatively faint dwarf galaxies is more efficient than implemented by the {\sc eagle}  galaxy formation model. Higher resolution simulations may also generate fewer galaxies. LA120, by contrast, fails to produce any systems that generate as many dwarf galaxies as observed.

 Inevitably this result depends on the galaxy formation efficiency of our model: if a sufficiently higher fraction of haloes is able to form galaxies than our galaxy formation model predicts, then the LA120 model will still be viable. To examine this point in more detail, we include in Fig.~\ref{LGLum} a plot of the galaxy abundance where haloes are selected by their $V_\rmn{max}$ rather than their stellar mass. We select haloes with \vmax$>15$\kms since this is the mass regime that contains many luminous and dark subhaloes. Two of the LA120 systems successfully exceed the observed galaxy abundance. Thus, the LA120 model remains viable, although disfavoured compared to CDM and LA10, for relatively bright ($M_{*}>10^{5}\Msun$) galaxies.

\begin{figure*}
    \includegraphics[scale=0.70]{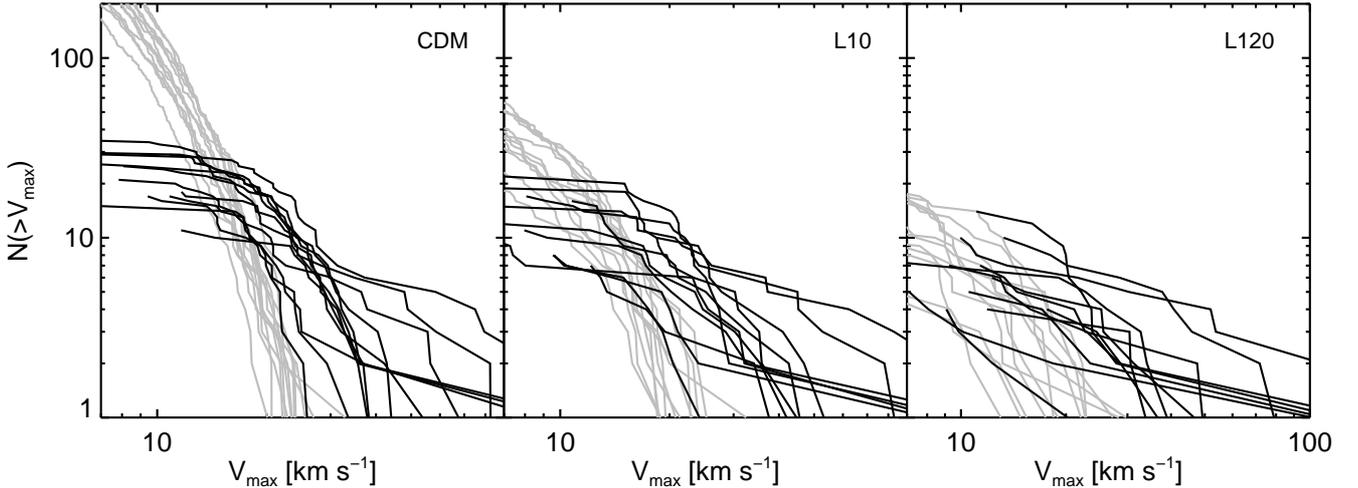}
    \caption{The distribution of luminous and dark substructures as a function of subhalo \vmax. Black lines denote luminous subhaloes and grey dark subhaloes. Left: CDM systems; centre: LA10 systems; right: LA120 systems.}
    \label{LaD}
  \end{figure*}

It is known from numerous studies using dark matter-only simulations that the dark matter halo mass function is strongly suppressed in WDM \citep{Bode01,Polisensky2011,Schneider13, Lovell14, Bose16b, Horiuchi16}. Would we therefore not expect to see a larger difference in the number of dark subhaloes? We show this comparison in Fig.~\ref{LaD}, where we plot the subhalo \vmax functions of MW and M31 satellites. We find that all subhaloes with \vmax$>33$~\kms are luminous in all of the dark matter models, where we define a (sub)halo to be luminous if it contains at least one star particle. Below that \vmax, the number of dark satellites starts to rise, and rapidly so in CDM \citep{Sawala16b}. Below $13$~\kms, the disparity between CDM and LA120 has become large enough that their distributions no longer overlap; LA10 similarly peels away below $10$~\kms although this will also be in part due to resolution effects. The detection of a large population of dark substructures, e.g. by means of lensing \citep{Vegetti14,Hezaveh16,LiR16} or stellar stream disruption \citep[e.g.][]{Carlberg16,Erkal16}, could rule out this sterile neutrino model. 

\begin{figure}
    \includegraphics[scale=0.33]{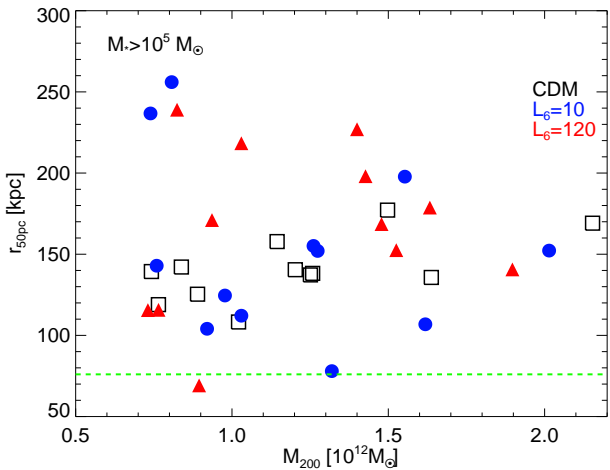}
    \caption{The radius that encloses 50~per~cent of $M_{*}>10^{5}\Msun$ as a function of $M_{200}$.  The black squares, blue circles and red triangles denote the CDM, LA10 and LA120 systems respectively. The dashed green line shows the value of $r_\rmn{50~p.c.}$ for the compiled list of \citet{McConnachie12}.}
    \label{RadDist}
  \end{figure}

  We conclude our discussion of satellite abundances with the radial distributions. The WDM haloes are less dense than their CDM counterparts (as discussed in Section~\ref{ss:struc}), and therefore the position of subhaloes around the main halo may differ due to dynamical friction and tidal stripping. In Fig.~\ref{RadDist} we plot the median distance to the main galaxy of satellites with $M_{*}>10^5\Msun$, which we denote $r_\rmn{50~p.c.}$, for CDM and our sterile neutrino models as a function of the host halo virial mass.  However, the sterile neutrino models are much more varied; this may also be a consequence of small number statistics in the smallest host haloes. All three models consistently predict median $r_\rmn{50~p.c.}$ larger than that measured for the MW satellite system. The median concentration is also related to the free-streaming length: 130~kpc for CDM, 150~kpc for LA10 and 170~kpc for LA120. However, the most concentrated system is a LA120 halo (albeit with only seven satellites) and the best match to the observed MW was simulated with the LA10 model.
   
\begin{figure*}
    \includegraphics[scale=0.70]{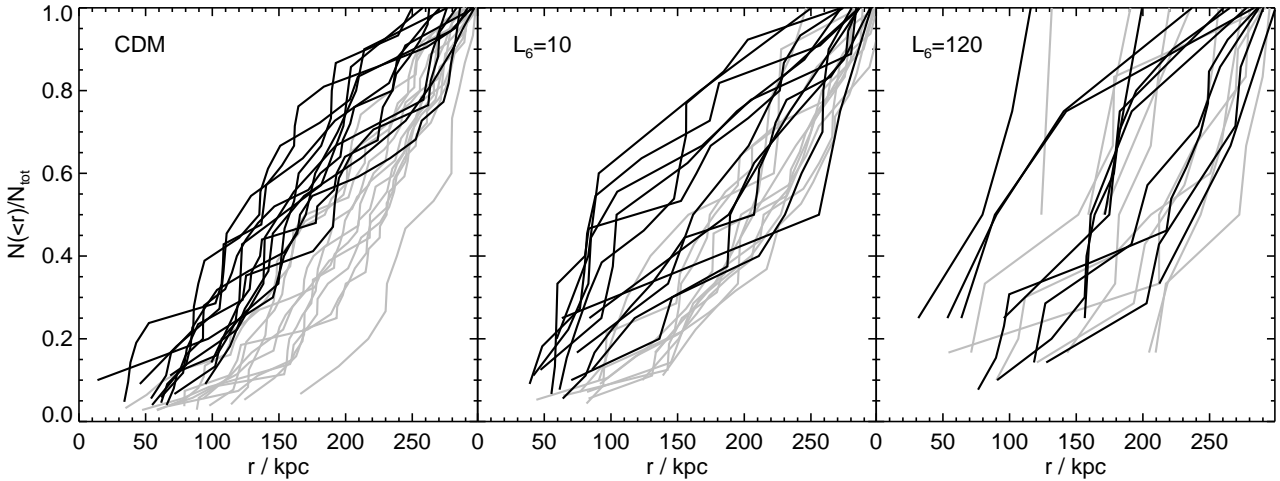}
    \caption{Cumulative radial distributions of subhaloes in CDM (left), LA10 (centre) and LA120 (right). Black lines denote luminous subhaloes (those that contain at least one star particle) and grey dark subhaloes. In both populations the minimum subhalo mass is $2\times10^{8}\Msun$.}
    \label{RadDistLD}
  \end{figure*}

The radial profiles are key to predicting the number of undiscovered satellites, helping us to estimate how many remain undetected due to their location being beyond the reach of current surveys. It has been shown that in the CDM cosmology luminous subhaloes are located closer to the centre of the halo than the subhalo population as a whole \citep{Wang13,Sawala16b}. To investigate whether this effect holds for our sterile neutrino models, in Fig.~\ref{RadDistLD} we plot the   cumulative distribution of luminous and dark satellites as a function of radius. We restrict our analysis to subhaloes with a total mass $>2\times10^{8}\Msun$ to retain only those subhaloes that have sufficient mass to form a galaxy. In the CDM case we recover the previously identified phenomenon that the locations of luminous subhaloes are more centrally concentrated than those that are dark: the median position of a luminous subhalo is $\approx140$~kpc whereas that of the dark subhaloes is $\approx50$~kpc further out. This separation is retained in LA10, albeit with greater scatter between the distributions due to increased shot noise.  No separation is apparent for LA120; however the Poisson error completely dominates in this case, both for the luminous and dark subhaloes. In summary, we conclude that the more concentrated distribution of visible satellites compared to dark satellites seen in CDM is also present in WDM models, at least in those that are compatible with observational constraints.
  
\subsection{Ages and metallicities}

\begin{figure}
    \includegraphics[scale=0.42]{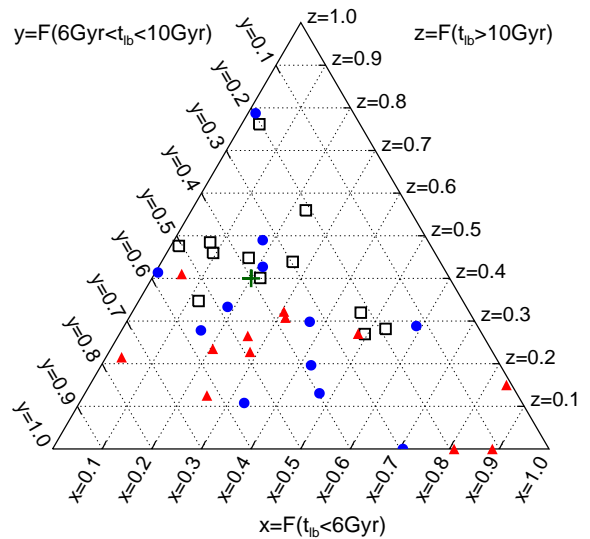}
    \caption{The relative abundances by age of stars in satellites in the stellar mass range $10^6<M_{*}/\Msun<10^7$. For each M31 and MW system we add together all of the satellites in the stellar mass bin such that there is one symbol per system. The $x$-axis shows the proportion of stars that are younger than 6~Gyr, the $z$-axis shows the proportion that are older than 10~Gyr, and the $y$-axis shows the proportion that are within this age range. The scheme for plotting the data points is the same as in Fig.~\ref{VmaxComp}. The approximate measured values of these quantities for the Local Group dwarf spheroidals as a whole, as presented by \citet{Weisz11}, are shown as the green cross.}
    \label{TP}
  \end{figure}

 The delay in WDM structure formation affects the ages of the satellites'  stellar populations so that WDM satellites have younger stellar populations than their CDM counterparts. This effect depends on the stellar mass of the satellite; the differences are likely to be largest for the smallest satellites, although in practice there will be stochastic deviations amongst the formation times of individual satellites. Some of the delay in formation times is also driven by resolution effects \citep{Starkenburg17}. We therefore select the star particles contained within satellites in the mass bin $\log_{10}(M_{*}/\Msun)=[6,7]$, and for each simulated system split the particles in to three bins in lookback time, denoted as $t_\rmn{lb}$: $t_\rmn{lb}<6$~Gyr, 6~Gyr~$\ge t_\rmn{lb}<10$~Gyr, and 10~Gyr~$\ge t_\rmn{lb}$. We plot the proportion of stellar mass in these bins in Fig.~\ref{TP} along with that of the measured Local Group dwarf spheroidals as compiled by \citet{Weisz11}.
  
 There are clear differences between the different dark matter models, albeit with a familiar, large overlap. At least 26~per~cent of the stars in all CDM systems are more than 10~Gyr old, whereas four LA10 systems and six LA120 systems do not meet this threshold. The largest proportion of $t_\rmn{lb}>10$~Gyr stars in L120A is 41~per~cent, younger than eight of the CDM systems. LA120 is instead biased towards more intermediate age systems, and LA10 systems take a large range of ages. The measured age makeup of the observed Local Group is located easily within the CDM and LA10 distributions but just outside the LA120. It is thus possible that the LA120 free-streaming length produces satellites that are too young compared to the Local Group, however much better resolution is required to confirm this conclusion.

\begin{figure}
    \includegraphics[scale=0.33]{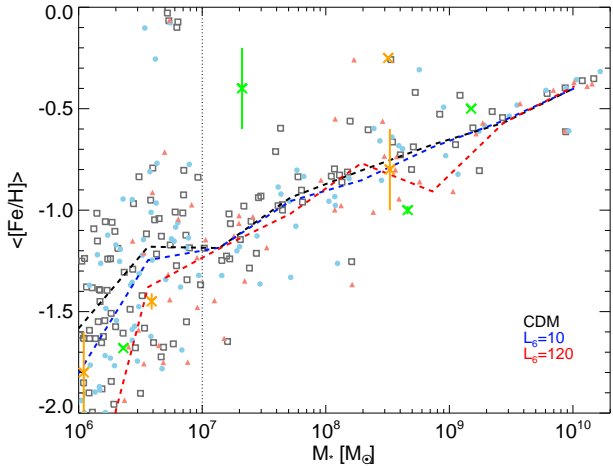}
    \caption{Stellar iron to hydrogen abundance as a function of stellar mass for satellite galaxies. Black symbols denote CDM galaxies, blue LA10 galaxies and red LA120 galaxies. The scheme for plotting the data points and medians is the same as in Fig.~\ref{VmaxComp}. Green and orange crosses denote the observed abundances for the MW and M31 satellites, respectively. The dotted black line marks the approximate 100 particle resolution limit.}
    \label{FeH}
  \end{figure}

 One proxy for the age of a galaxy is its metallicity, in that metal poorer stars form in gas that has been less enriched by prior star formation events. Observationally, the  iron-to-hydrogen abundance ratio, $\langle[\rmn{Fe/H}\rangle]$, has been measured for most Local Group galaxies. We compare these observations as a function of their stellar masses to our simulations results in Fig.~\ref{FeH} . For all three models we recover the result from the {\sc eagle} simulations that fainter satellites have lower stellar metallicities, as one would expect from having both formed earlier and undergone a lower recycling rate. The LA10 and LA120 metallicities are suppressed slightly relative to CDM at a given mass, at least in the range   $M_{*}<4\times10^{8}\Msun$, and the former are in better agreement with the observational data. Such a result should be treated with caution, however, as there remain large systematics in both the observations (e.g. limited number of stellar spectra calibration) and simulations (sensitivity to the feedback implementation, nucleosynthesis yields, type Ia supernova rates, see  \mbox{\citealp{Schaye15}} for details  \mbox{\citealp[c.f.][]{Onorbe15}}).   
  
  \section{Conclusions}
  \label{Con}

Our motivation to undertake this study was to examine the properties of Local Group-analogue systems in which the dark matter is made of sterile neutrinos rather than WIMPs. Sterile neutrino models resolve several particle physics problems, and may show an imprint on the properties of satellite galaxies. The sterile neutrino is relativistic when it decouples from the primordial plasma and can thus free stream out of potential wells corresponding to the scales of dwarf galaxies. The free-streaming length of the sterile neutrinos is a function of the sterile neutrino mass and mixing angle, both of which also determine the frequency and amplitude of the sterile neutrino decay into X-ray photons. Candidate decay signals have been proposed as a source of the unexplained 3.5~keV line feature in X-ray spectra.   \mbox{\citep[e.g.][]{Boyarsky14a,Bulbul14}}. Such a sterile neutrino would have a mass of 7.1~keV and a mixing angle of $\sim10^{-11}$. We thus investigate the properties of satellite galaxies for which the underlying dark matter consists of 7~keV sterile neutrinos rather than CDM, and determine whether this sort of WDM model is more in keeping with the observations of small galaxies compared to predictions from CDM simulations.
  
 We thus resimulated six Local Group-analogue systems from the APOSTLE project with two 7~keV sterile neutrino models. One of these was chosen to be consistent with the unexplained 3.5keV line detected in clusters and M31 (LA10), and the second to have the longest free-streaming length of any 7~keV sterile neutrino (LA120). We showed that these models preserved the large-scale structure of the Local Group systems and therefore maintain the successes enjoyed by the \LCDM~ paradigm on LG scales.
  
 We first considered the structure of the satellites. We showed that the relationship between the satellite \vmax and stellar mass was very similar, if not identical, between CDM and the sterile neutrino models (Fig.~\ref{MsubMstar}). We determined that the concentrations of LA10 subhaloes are lower than those of CDM and that those of LA120 are lower still (Fig.~\ref{VmRm}). This results in stronger stripping of the sterile neutrino satellites and, hence, higher baryon fractions than for the CDM satellites. We also considered the effect of baryon physics, feedback from reionization and supernovae, on the halo structure and found that all three models exhibited the same behaviour, namely a suppression of 12~per cent in their \vmax values (Fig.~\ref{VmaxComp}) and 20~per~cent in their densities  within 1~kpc relative to what one would expect from a dark matter-only simulation (Fig.~\ref{M2DMO}). We also showed that satellites with the stellar mass and central density of the Fornax dwarf spheroidal are closer to the median of the stellar mass-central density distribution in the sterile neutrino models than in CDM, and thus a clear way to discriminate between CDM and sterile neutrino models is to find evidence for a population of faint ($M_{*}<10^{6}\Msun$) yet dense ($V_\rmn{1kpc}>20$)~\kms satellites. This requires a concerted effort to obtain high quality spectra of stars in the known, faint satellites from which the velocity dispersion can be measured.

  The number density of satellite galaxies was found to be suppressed in the LA10 systems by a median of 45 per cent compared to CDM. However, the stochastic scatter between systems blurs the distinction between the two: from the satellite luminosity function alone it is not possible to tell whether a given system is a satellite-poor CDM or a satellite-rich LA10 version (Fig.~\ref{SatLum}). The suppression is stronger for LA120 to the point that this model fails to reproduce the total number of Local Group galaxies. This model is ruled out by our simulations unless: (i) galaxy formation in small haloes is more efficient than is prescribed by the {\sc eagle} model, or (ii) the Local Group is much more massive than assumed in APOSTLE. Higher resolution simulations are required to extend this analysis to fainter galaxies, the regime where the viability of LA120 is more questionable  \mbox{\citep[c.f.][]{Lovell16,Schneider16}.}

  There is an expectation that, since the formation of the lowest mass satellites is delayed, the star formation histories and metallicities of dwarf galaxies will change. We found evidence for systematically delayed formation times only in dwarf galaxies with masses with $M_{*}<10^7\Msun$ (Fig.~\ref{TP}). There is some evidence for lower $\langle[\rmn{Fe/H}]\rangle$ abundances for both sterile neutrino models; however, the scatter is again dominant and so a much bigger sample of satellites with better mass resolution would be required to make a quantitative statement.
 
 We have shown in this work that the LA10 model is sufficiently close to CDM in its predicted satellite properties that the 3.5keV line-consistent 7~keV sterile neutrino is a viable alternative to the WIMP, whereas the results for the LA120 model demonstrate that many sterile neutrino models are in tension with current observations. Inevitably, however, the baryon physics, as stated above, cannot be modelled from first principles and so this introduces a systematic uncertainty into our model. Therefore, the best means for differentiating between CDM and sterile neutrino models rely on identifying dark subhaloes, which at $10$\kms  show clearly different abundances between these three models. Methods for detecting substructure in lensing arcs and gaps in streams will therefore hold the key, from an astrophysical perspective, to determine which class of particle physics candidate will be identified as the dark matter  \mbox{\citep{LiR16}}. Possible confusion with baryonic substructures will need to be accounted for  \mbox{\citep[e.g.][]{Amorisco16}}.

  \section*{Acknowledgements}
	MRL would like to thank Vasily Belokurov for useful discussions. We would like to thank Mikko Laine for supplying the code that calculates the sterile neutrino distribution functions. This work was carried out on the Dutch National e-Infrastructure with the support of SURF Cooperative, and also the DiRAC Data Centric system at Durham University, operated by the Institute for Computational Cosmology on behalf of the STFC DiRAC HPC Facility (www.dirac.ac.uk). This equipment was funded by BIS National e-Infrastructure capital grant ST/K00042X/1, STFC capital grant ST/H008519/1, and STFC DiRAC Operations grant ST/K003267/1 and Durham University. DiRAC is part of the National e-Infrastructure. This work is part of the D-ITP consortium, a programme of the Netherlands Organization for Scientific Research (NWO) that is funded by the Dutch Ministry of Education, Culture and Science (OCW).  This work was supported in part by an STFC
 rolling grant to the ICC and by ERC Advanced Investigator grant COSMIWAY [GA 267291] and starting grant 278594-GasAroundGalaxies, and also the Interuniversity Attraction Poles Programme of the Belgian Science Policy Office [AP P7/08 CHARM]. RAC is a Royal Society University Research Fellow. ADL is supported by a COFUND Junior Research Fellowship. WAH acknowledges support from European Research Council (grant number 646702 `CosTesGrav') and the Polish National Science Center under contract no.~UMO-2012/07/D/ST9/02785. TS acknowledges the support of the Academy of Finland grant 1274931. This project has received funding from the European Research Council (ERC) under the European Union's Horizon 2020 research and innovation programme (GA N$^{\circ}$ 694896).

  \bibliographystyle{mnras}

\appendix
\section{Spurious subhaloes}
\label{app:ss}

\begin{figure}
    \includegraphics[scale=0.5]{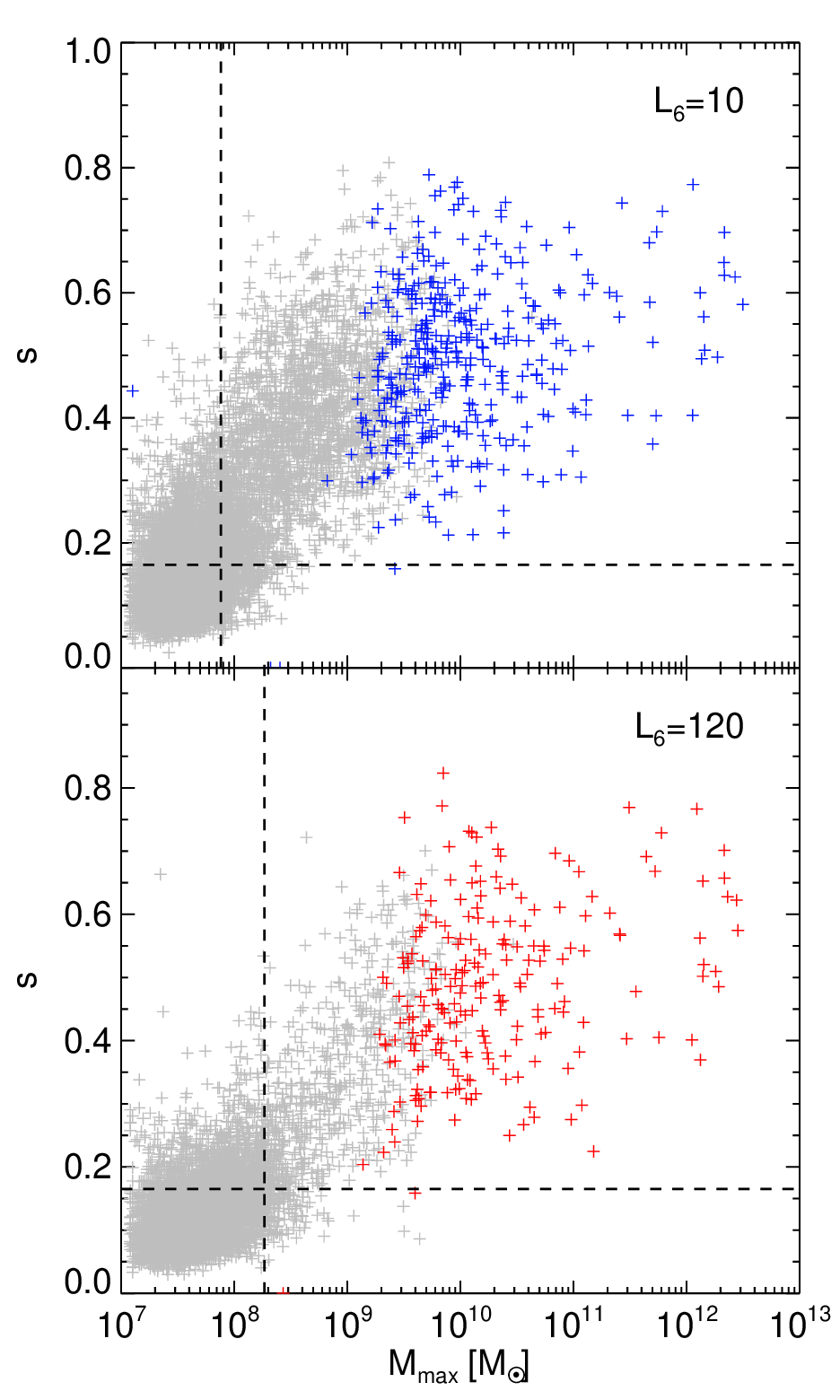}
    \caption{Sphericity of the initial conditions Lagrangian region sphericity, $s$, of each halo vs. maximum halo mass.  Top: results for LA10, bottom: LA120. Haloes that do not contain any star particles are shown in grey, those that contain at least one star particle are shown in colour: red for LA120 and blue for LA10. The vertical dashed lines denote the maximum mass criteria derived in \citet{Lovell14}, and the horizontal line the sphericity threshold $s=0.165$, below which haloes are deemed to be spurious. Only subhaloes in the top right parts of the diagrams are considered in this work. None of the possibly spurious subhaloes contains stars.}
    \label{MMvSpher}
  \end{figure}

One additional challenge to simulating systems with WDM, or indeed any model with a steep power spectrum cutoff, is that filaments fragment artificially into small, bound structures that dominate the halo mass function below some characteristic mass \citep{Wang07,Lovell14,Power16}. If galaxies were to form in these spurious haloes our estimates for galaxy  abundances in the sterile neutrino model would be compromised. To check for evidence of any issues with spurious haloes, we employ the method of \citet{Lovell14}. We calculate the sphericity of the Lagrangian regions in the initial conditions of each of our haloes: those haloes with low sphericities will have collapsed from slices of filaments and are therefore spurious. The precise method requires the sphericity calculation to be performed using the particles that the haloes had when they first formed, i.e. the time at which their mass was half that of the maximum mass attained during the simulation. We plot this half-maximum mass sphericity against the maximum mass  -- a complementary spurious subhalo diagnostic -- in Fig.~\ref{MMvSpher}. This figure includes all haloes and subhaloes that are within 2~Mpc of the MW-M31 barycentre across all six of our volumes for both LA10 and LA120. 

  The trend observed in \citet{Lovell14} that small, spurious haloes have low sphericities is replicated in our hydrodynamical runs below a maximum mass of $\sim3\times10^8\Msun$ in LA10 and $\sim1\times10^9\Msun$ in LA120. Crucially, those haloes that host even one star particle  have maximum masses above $10^9\Msun$. Therefore, we are confident that spurious haloes do not host galaxies. Indeed, there are many haloes above $10^9\Msun$ that do not form or host galaxies and will therefore be dark subhaloes (see Fig.~\ref{LaD} and the accompanying discussion). In the remainder of the paper, where we consider dark subhaloes, we remove haloes from our catalogues using the half-maximum sphericity, $s$, and maximum mass, $M_\rmn{max}$ criteria of \citet{Lovell14}. The maximum mass cuts are  $M_\rmn{max}=7.6\times10^{7}\Msun$ for LA10 and $M_\rmn{max}=1.9\times10^{8}\Msun$ for LA120 as shown by dashed lines in Fig.~\ref{MMvSpher}; the sphericity cut is $s=0.165$ for both models. Note that $M_\rmn{max}$ is derived as a function of the peak wavenumber of the power spectrum: we adopt $k_\rmn{peak}=9.4~h~\rmn{Mpc}^{-1}$ for LA10 and $k_\rmn{peak}=6.0~h~\rmn{Mpc}^{-1}$ for LA120. We cannot rule out that dark, spurious subhaloes are able to influence the process of galaxy formation within genuine haloes, but we expect any such effects to be minor since spurious subhaloes contain only a small fraction of the mass.

  \end{document}